\documentclass[final,5p,times,twocolumn]{elsarticle}

\usepackage{amsfonts,amssymb,amsmath,graphicx,amscd,ulem}
\usepackage{array}
\usepackage[latin1]{inputenc}

\def\bi{\begin{itemize}}

\def\ei{\end{itemize}}
\def\be{\begin{equation}}
\def\ee{\end{equation}}
\def\bea{\begin{eqnarray}}
\def\eea{\end{eqnarray}}

\def\gdot{\dot\gamma}

\def\upd{\text{d}}

\def\m1{$^{-1}$}

\DeclareTextSymbol{\degre}{OT1}{23}

\journal{Journal of Non-Newtonian Fluid Mechanics}

\begin{document}

\begin{frontmatter}



\title{On the existence of a simple yield stress fluid behavior}


\author[adr1]{G. Ovarlez\corref{cor1}}
\author[adr2,adr3]{S. Cohen-Addad}
\author[adr4]{K. Krishan}
\author[adr1]{J. Goyon}
\author[adr1]{P. Coussot}
\address[adr1]{Université Paris Est, Laboratoire Navier (UMR CNRS 8205), 2 allée Kepler, 77420 Champs-sur-Marne, France}
\address[adr2]{Univ. Paris 6, UMR 7588 CNRS-UPMC, INSP, 4 place Jussieu, 75252 Paris cedex 05}
\address[adr3]{Université Paris-Est, LPMDI,
5 Bd Descartes, 77454 Marne-la-Vallée, France}
\address[adr4]{P\&G Kobe Technical Center,
1-17 Koyocho Naka, Higashinada-ku, Kobe, 6580032 Hyogo, Japan}
\cortext[cor1]{corresponding author: guillaume.ovarlez@ifsttar.fr}

\begin{abstract}
Materials such as foams, concentrated emulsions, dense suspensions
or colloidal gels, are yield stress fluids. Their steady flow
behavior, characterized by standard rheometric techniques, is
usually modeled by a Herschel-Bulkley law. The emergence of
techniques that allow the measurement of their local flow
properties (velocity and volume fraction fields) has led to
observe new complex behaviors. It was shown that many of these
materials exhibit shear banding in a homogeneous shear stress
field, which cannot be accounted for by the standard steady-state
constitutive laws of simple yield stress fluids. In some cases, it
was also observed that the velocity fields under various
conditions cannot be modeled with a single constitutive law and
that nonlocal models are needed to describe the flows. Doubt may
then be cast on any macroscopic characterization of such systems,
and one may wonder if any material behaves in some conditions as a
Herschel-Bulkley material. In this paper, we address the question
of the existence of a simple yield stress fluid behavior. We first
review experimental results from the literature and we point out
the main factors (physical properties, experimental procedure) at
the origin of flow inhomogeneities and nonlocal effects. It leads
us to propose a well-defined procedure to ensure that steady-state
bulk properties of the materials are studied. We use this
procedure to investigate yield stress fluid flows with MRI
techniques. We focus on nonthixotropic dense suspensions of soft
particles (foams, concentrated emulsions, Carbopol gels). We
show that, as long as they are studied in a wide (as
compared to the size of the material mesoscopic elements) gap
geometry, these materials behave as `simple yield stress fluids':
they are homogeneous, they do not exhibit steady-state shear
banding, and their steady flow behavior in simple shear can be
modeled by a local continuous monotonic constitutive equation which accounts for
flows in various conditions and matches the macroscopic response.

\end{abstract}


\end{frontmatter}

\section{Introduction}\label{section_introduction}

Materials such as dense suspensions, colloidal gels, microgel
suspensions, concentrated emulsions or foams, are yield stress
fluids: below a yield stress $\tau_y$, they do not
flow\footnote{We neglect slow creep flows driven by coarsening
dynamics in foams and emulsions\cite{Hohler2005}.}. Their steady
flow behavior in simple shear is characterized in standard
rheological experiments, e.g. in a Couette or a cone-and-plate
geometry \cite{Macosko}. It is usually well accounted for by a
Herschel-Bulkley equation
$\tau(\gdot)=\tau_y+\eta_{_{HB}}\gdot^n$, the parameters of which
depend on the details of the material microstructure
\cite{Coussot2005}. The ability of such models to help predicting
flows in more complex configurations, e.g. in extrusion flows
\cite{Rabideau2010}, then relies on the robustness of the
characterization in simple shear.

However, the recent emergence of techniques
\cite{Coussot2005,Manneville2008} that allow the measurement of
local flow properties (velocity and volume fraction fields) has
led to revisit the behavior of pasty materials. The reason is
that, in rheometric experiments, one only measures macroscopic
quantities (torque $T(\Omega)$ vs. rotational velocity $\Omega$).
The material constitutive law relating local quantities (shear
stress $\tau(\gdot)$ vs. shear rate $\gdot$) is then derived under
the assumption that the flow and the material are homogeneous, and
that the material can be modeled as a continuum with a local
constitutive law. Local measurements have shown that these
conditions are actually not met for some materials. The
observations of steady flow inhomogeneities (shear banding) in
homogeneous stress fields have in particular shown that, in
addition to their yield stress $\tau_y$, some materials are also
characterized by a critical shear rate $\gdot_c$ below which they
cannot flow steadily \cite{Dennin2008,Ovarlez2009,Schall2010}.
Moreover, in some materials, it was found that flows under various
conditions cannot be described by a single local constitutive law
\cite{Goyon2008,Katgert2010}, which led to propose nonlocal
modeling of their behavior \cite{Goyon2008,bocquet2009}. These
observations contrast with standard laws inferred from purely
macroscopic measurements, such as that of Herschel-Bulkley, which
are local and allow steady flow at any low shear rate $\gdot$.

Doubt may finally \textit{a priori} be cast on any macroscopic
characterization of yield stress fluids, and one may wonder if any
material behaves in some conditions as a Herschel-Bulkley
material. The question we address in this paper is whether such a
simple yield stress fluid behavior exists or not. In
Sec.~\ref{section_sh_bdg}, we first focus on the shear banding
issue; then, in Sec.~\ref{section_local}, we investigate the
consistency between yield stress fluids flows and a local modeling
of their behavior. Both sections are structured in the same way.
We first review experimental studies of yield stress fluid flows
in the literature: we present the main sources of shear banding
(Sec.~\ref{subsection_intro_shbdg}) and of discrepancy between
macroscopic and local behavior
(Sec.~\ref{subsection_local_review}). This leads us to distinguish
three classes of materials: thixotropic materials (such as
colloidal gels), nonthixotropic dense suspensions of rigid
particles (such as noncolloidal suspensions and colloidal
glasses), and nonthixotropic dense suspensions of soft particles
(such as foams, concentrated emulsions and Carbopol gels). We
discuss the role of thixotropy and shear-induced migration in the
studied complex behaviors for the first two classes of materials.
We then focus on nonthixotropic dense suspensions of soft
particles (Secs.~\ref{subsection_soft}
\&~\ref{subsection_local_review}), whose case is yet unclear as
contradictory observations are reported in the literature. We
first show that the above mechanisms should not be at play in
these materials. We then point out issues related to the procedure
used to study their flows. We propose a well-defined procedure to
ensure that steady-state bulk properties of the materials are
studied, and we investigate the flows of several foams,
concentrated emulsions, and a Carbopol gel, with MRI techniques
(Secs.~\ref{subsection_soft_noshbdg}
\&~\ref{subsection_local_soft}). We show that, as long as they are
studied in a wide (as compared to the size of the material
mesoscopic elements) gap geometry, these materials behave as
`simple yield stress fluids': they are homogeneous, they do not
exhibit steady-state shear banding, and their steady flow behavior
in simple shear can be modeled by a local continuous monotonic
constitutive equation which accounts for their flows in various
conditions and matches their macroscopic response.

\section{Shear banding}\label{section_sh_bdg}

In this section, we address the question of shear banding in yield
stress fluids. We discuss the phenomenon and the origins
identified in two classes of such materials: structural
inhomogeneities in thixotropic systems; volume fraction
inhomogeneities in dense suspensions of rigid particles. We then
investigate flow stability in nonthixotropic dense suspensions of
soft particles. We discuss the relevance of the above mechanisms
of shear banding in these systems. We use a procedure designed to
circumvent possible artefacts and show that these materials do not
display any steady-state shear banding, in contrast with several
results of the literature. Possible explanations of these
discrepancies are proposed at the end of the section.

\subsection{Phenomenon and possible origins}\label{subsection_intro_shbdg}

Shear banding in yield stress fluid flows here refers to the
coexistence of liquid-like (sheared) and solid-like (unsheared)
bands at steady-state in a homogeneous shear stress
field\footnote{It is also sometimes called `discontinuous' shear
banding to avoid confusion with other kinds of flow
inhomogeneities.}. It is observed in some materials when a
macroscopic shear rate $\gdot_{\text{macro}}$ lower than a
critical value $\gdot_c$, which depends on the material, is
applied. In these conditions, the material basically splits into a
region flowing at $\gdot_c$ and a non flowing region, whose
relative extent ensures that the shear rate spatial average equals
$\gdot_{\text{macro}}$ \cite{Coussot2002a,Moller2008,Ovarlez2009}.
The critical shear rate $\gdot_c$ is the shear rate below which no
stable flow is possible. It appears to be a new mechanical
characteristics of these systems, in addition to their yield
stress $\tau_y$: at steady state, flow implies both
$\tau\geq\tau_y$ and $\gdot\geq\gdot_c$. In an inhomogeneous
stress field, flow localization occurs with any yield stress
fluid; the signature of shear banding is then that the material
flows at a nonzero shear rate $\gdot_c$ at the interface between
the flowing region (where the shear stress $\tau$ is higher than
$\tau_y$) and the non-flowing region (where $\tau<\tau_y$)
\cite{Ovarlez2009}. This phenomenon is generally attributed to the
fact that the theoretical material steady-state flow curve is
non-monotonic \cite{Olmsted2008}.

\begin{figure}[htb]\begin{center}
\includegraphics[width=8.5cm]{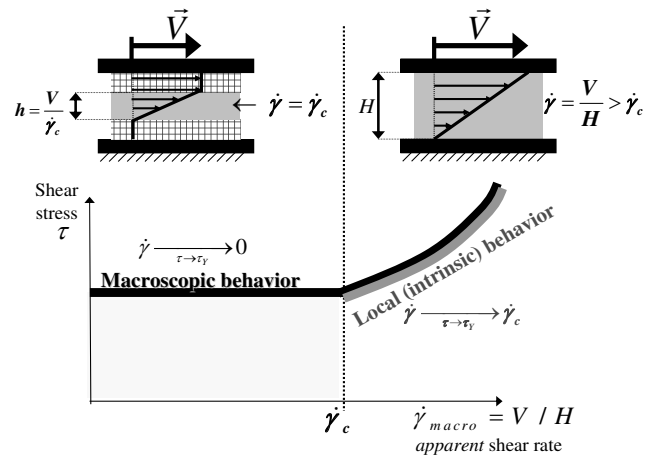}
\caption{Illustration of the difference between the material
intrinsic constitutive law (grey line) and the flow curve measured
macroscopically in a homogeneous stress field (black line) in a
shear-banding material. At high macroscopic shear rate
($\gdot_{\text{macro}}>\gdot_c$), flow is homogeneous and the
macroscopic flow curve matches the local behavior. At low
macroscopic shear rate ($\gdot_{\text{macro}}<\gdot_c$), shear
banding occurs: the local shear rate in the flowing material --
$\gdot_c$ -- differs from the macroscopic shear rate and the
macroscopic flow curve does not match the local
behavior.}\label{figure_sketch}
\end{center}\end{figure}

Although shear-banding materials cannot flow steadily at a local
shear rate smaller than $\gdot_c$, any macroscopic shear rate
$\gdot_{\text{macro}}<\gdot_c$ can be applied to such system by
the relative motion of two boundaries. This implies that their
macroscopic characterization may fail to represent their actual
behavior. Indeed, typical constitutive laws
$\tau(\gdot_{\text{macro}})$ found near the yield stress in
macroscopic experiments are such that $\gdot_{\text{macro}}$ tends
continuously to zero as $\tau$ approaches $\tau_y$; this is the
case of standard laws such as the Herschel-Bulkley law. This
contrasts with the locally observed behavior of shear-banding
materials where $\gdot_{local}$ tends towards a critical shear
rate $\gdot_c\neq0$ as $\tau$ approaches $\tau_y$. This issue is
illustrated in Fig.~\ref{figure_sketch}.

Identifying which materials exhibit shear bands and which do not
is thus of high importance; only the last ones may behave like
simple yield stress fluids and be characterized using standard
rheometric experiments. Two possible origins of shear-banding in
yield stress fluids have already been identified experimentally:
structural inhomogeneities in thixotropic systems, and volume
fraction inhomogeneities in dense suspensions of rigid particles.

\subsubsection{Structural inhomogeneities in thixotropic systems}\label{subsubsection_thixotropy}

Shear banding in yield stress fluids seems to be observed in all
thixotropic colloidal gels
\cite{Coussot2002a,Ragouilliaux2007,Moller2008,Rogers2008,Ovarlez2009}.
In these materials, it is attributed to the competition between
aggregation, which is due to attractive interactions and thermally
activated structuration, and shear, which tends to break the
aggregates. Starting from a liquid state, when the applied shear
rate $\gdot$ is too low, the material basically splits between a
liquid-like region (where it is a suspension of independent
aggregates) and a region that gets continuously structured under
shear until it jams (i.e., a percolated network forms). Shear
banding is also observed in dense suspensions of rigid
noncolloidal particles, where it is attributed to competition
between sedimentation-induced contact formation and shear-induced
resuspension \cite{Ovarlez2006,Fall2009}. When the local shear
rate is too low, the shear resuspension mechanism is not
sufficiently efficient to prevent contact formation between the
particles and the material jams as a percolated network of
particles form.

In these two kinds of systems, shear banding seems to result from
shear\!\!~/\!\! structure coupling, and from the development of
two bands of different structures, similarly to what is observed
in wormlike micellar systems \cite{berret1997}. Several models
based on the idea that the material properties depend on its
structure and that the structure depends on shear history
\cite{Coussot2002b,Fielding2009,Mansard2011} have indeed predicted
shear banding in simple shear, as a result of the instability of
some solutions of homogeneous structure, and describe correctly
the flow inhomogeneities observed in thixotropic systems
\cite{Coussot2005}. Such models can be an explanation for shear
banding in a given material only if a macroscopic thixotropic
behavior -- i.e., dependent on shear history \cite{Mewis2008} --
is found experimentally.

\subsubsection{Volume fraction inhomogeneities in suspensions of rigid particles}\label{subsubsection_migration}

Colloidal glasses are not much thixotropic: they get structured at
rest and destructured at flow start-up, but their flow behavior
does not depend significantly on their flow history (see e.g.
\cite{Derec2003}). The above mechanism is thus not expected to be
at play in such systems. However, shear banding has recently been
observed in these materials and explained to result from the
development of an inhomogeneous volume fraction profile
\cite{besseling2010}. The basic idea is that a small change in the
volume fraction may result in a significant change in the
rheological properties, in particular in the yield stress value,
which is an increasing function of $\phi$. In a homogeneous stress
field, at low applied shear rate, regions of lower $\phi$ (and
thus of lower $\tau_y$) may then flow while regions of higher
$\phi$ (and thus of higher $\tau_y$) are jammed. The development
of an inhomogeneous volume fraction profile in a homogeneous
stress field suggests an instability induced by a
shear-concentration coupling \cite{Schmitt1995,besseling2010}.
Banding resulting from volume fraction inhomogeneities has also
been observed in dense rigid noncolloidal suspensions
\cite{Ovarlez2006,Fall2010}, where shear-induced migration
\cite{Phillips1992,Morris1999,Lhuillier2009} creates jammed
regions of volume fraction $\phi$ higher than the jamming packing
fraction $\phi_m$.

It is worth noting that these systems are made up of rigid
particles. Indeed, although hydrodynamic interactions may play a
role, recent theoretical developments \cite{Lhuillier2009}
supported by experimental findings \cite{Fall2010} show that the
shear-induced migration efficiency is strongly enhanced by direct
rigid contact forces between the particles. This characteristic
may thus play a crucial role in the emergence of shear banding in
these systems.

\subsection{Nonthixotropic dense suspensions of soft particles}\label{subsection_soft}

The case of nonthixotropic dense suspensions of soft particles
(foams, concentrated emulsions, microgels) is less clear, as
contradictory observations are reported in the literature.
\citet{becu2006} reported shear banding in adhesive emulsions,
whereas \citet{Ovarlez2008} did not observe any shear banding in
the same system as well as in four other adhesive and nonadhesive
emulsions. \citet{Rodts2005} reported shear bands in 3D sheared
foams, but \citet{Ovarlez2010b} did not observe any shear banding
in the same system and in three other foams. In 2D foams (bubble
rafts) \citet{Gilbreth2006} reported shear banding whereas
\citet{Katgert2010} did not. This suggests that subtle differences
in the materials, setup, or procedures, may exist, and have to be
identified. In the following, we first discuss the possible
relevance of the mechanisms mentioned in
Sec.~\ref{subsection_intro_shbdg} in these systems.

\subsubsection{Nonthixotropic behavior}\label{subsubsection_soft_structure}

\begin{figure*}[htb]
\begin{center}
\includegraphics[width=14cm]{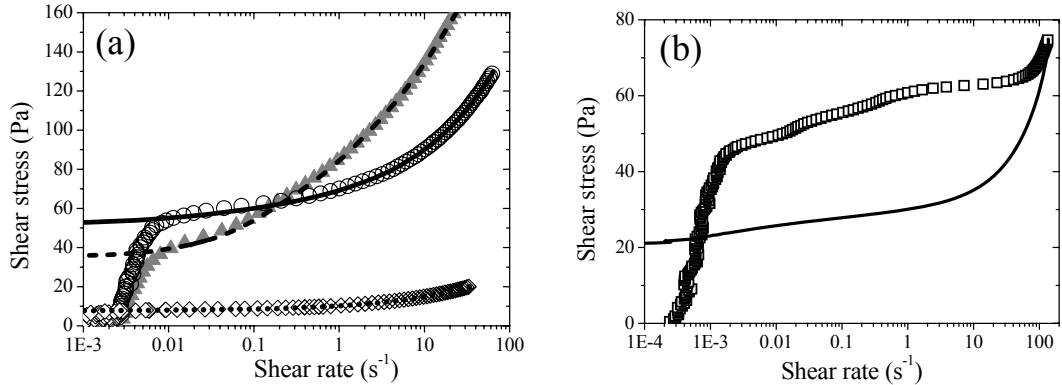}
\caption{Shear stress $\tau$ vs. shear rate $\gdot$ when
consecutive up (symbols) and down (lines) stress ramps are
applied, with a constant rate $\alpha$, (a) to a foam (up
triangles and dashed line, $\alpha$=1~Pa/s), a Carbopol gel (empty
circles and full line, $\alpha$=0.5~Pa/s) and a concentrated
emulsion (empty diamonds and dotted line, $\alpha$=0.1~Pa/s), and
(b) to a thixotropic (bentonite) suspension ($\alpha$=0.5~Pa/s). A
4\degre angle cone-and-plate geometry is used, with serrated
surfaces (60~mm diameter) for the foam, with sandblasted surfaces
(40~mm diameter) for the other materials. The emulsion is composed
of 1~$\mu$m diameter dodecane droplets dispersed at a 73\% volume
fraction in water stabilized by Sodium Dodecyl Sulfate at a 1 wt\%
concentration (see \cite{Ovarlez2010} for details). The Carbopol
gel is made up of Carbopol 980 dispersed and neutralized at a
0.3\% volume fraction in water (see \cite{Mahaut2008} for
details). The foam (Gillette Regular) consists of 27~$\mu$m
diameter bubbles at a 92\% volume fraction. The bentonite
suspension is made up of (smectite) clay particles of length of
order 1~$\mu$m and thickness 10~nm suspended at a 6\% volume
fraction in water \cite{Coussot2010}.}\label{fig_non_hysteretic}
\end{center}
\end{figure*}

The systems we deal with are basically nonthixotropic: their flow
behavior does not show significant history dependence. This is
illustrated in Fig.~\ref{fig_non_hysteretic}a, where the response
of an emulsion, a foam, and a Carbopol gel to consecutive up and down
stress ramps is displayed. Such experiment is classically used to assess the
thixotropic behavior of a material \cite{Coussot2005}. During the up-ramp, as long as the shear stress is below the
(static) yield stress, the material is strained in its solid
regime; this explains why the strain rate is roughly constant and
proportional to the ramp rate in a first stage \cite{Coussot2005}.
Once the material flows, there is no
significant difference between the responses to the up- and down-ramps: these materials can be
considered as nonthixotropic. This contrasts with the hysteretic
response observed in a thixotropic material (bentonite suspension, Fig.~\ref{fig_non_hysteretic}b), which is the signature of strong
dependence of the behavior on flow history, i.e., of thixotropy.

This suggests that the mechanisms discussed in
Sec.~\ref{subsubsection_thixotropy} should not be at play in the
Carbopol gel, the foam and the emulsion, as their structure should
remain homogeneous when they flow. Structuration at rest is nevertheless
observed in some of these systems
\cite{Hohler1999,Hohler2005,Divoux2011}, i.e., their structure in
their solid and liquid regimes may differ. E.g., after a preshear, foams progressively recover their elasticity at rest ($G'$ increases in time) as coarsening induces structural rearrangements \cite{Hohler1999,Hohler2005}. Structuration at rest can lead to differences between the static and dynamic
yield stress \cite{Divoux2011}. As detailed in
Sec.~\ref{subsubsection_soft_artefacts}, this can lead to
transient flow inhomogeneities but not to steady-state bands.

Recently, models of soft-jammed systems such as concentrated
emulsions (a nonlocal model \cite{Mansard2011} and the SGR model
\cite{Fielding2009}) have been extended to account for possible
bulk shear banding. Both models predict macroscopic shear banding
associated with structural inhomogeneities. We stress that, in
these models, the behavior depends on internal variables that
depend on shear history, i.e., the system is supposed to be
thixotropic: they cannot apply to \textit{nonthixotropic}
suspensions of soft particles.

\subsubsection{Material homogeneity}\label{subsubsection_soft_concentration}

As discussed in Sec.~\ref{subsubsection_migration}, volume
fraction inhomogeneities may lead to shear banding. One may wonder whether they can be responsible for some of the reported shear bands in dense suspensions of soft particles. Shear-induced migration is well documented
for rigid particles
\cite{Phillips1992,Morris1999,Ovarlez2006,Lhuillier2009,Fall2010}. Migration of deformable
particles, and particularly drops, has been much less studied. It
is found that single droplets migrate away from rigid walls, due
to asymmetric flows around deformed droplets
\cite{hollingsworth2006,karnis1967}, but binary collisions tend to
homogenize the systems for $\phi\geq 10\%$.

To test the possible emergence of volume fraction
inhomogeneities in dense suspensions of soft particles, we study the evolution in time of the bubble or droplet volume
fraction profiles when foams or concentrated emulsions are sheared in an inhomogeneous stress field (where particles migrate towards the low shear zones in suspensions of hard spheres). We use a wide gap Couette geometry, where the shear stress
distribution is: \bea\tau(R)=\tau(R_i)R_i^2/
R^2\label{eq_stress}\eea where $R$ is the radial position in the
gap and $R_i$ is the inner cylinder radius. The materials and methods are
described in detail in \citet{Ovarlez2008,Ovarlez2010b}. The
average bubble diameter of the studied foam is 73~$\mu$m, its gas
volume fraction is 89\% \cite{Ovarlez2010b}. Five emulsions are
studied; their droplet diameter and droplet volume fraction are:
(0.3~$\mu$m,75\%), (1~$\mu$m,75\%), (6.5~$\mu$m,75\%) and
(40~$\mu$m,88\%); both adhesive and nonadhesive emulsions are
studied \cite{Ovarlez2008}. The inner cylinder radius of the
Couette cell is 4.1~cm; the gap is 1.9~cm wide. The systems are
sheared at various constant rotational velocities $\Omega$,
ranging from 1 to 100 rpm, during long times (from 1 to 24~h,
depending on the system). The Couette cell is inserted in a
Magnetic Resonance Imager \cite{Rodts2004} and droplet or
bubble volume fraction profiles are measured as in
\cite{Ovarlez2006,Ovarlez2008}.

\begin{figure}[htb]\begin{center}
\includegraphics[width=7cm]{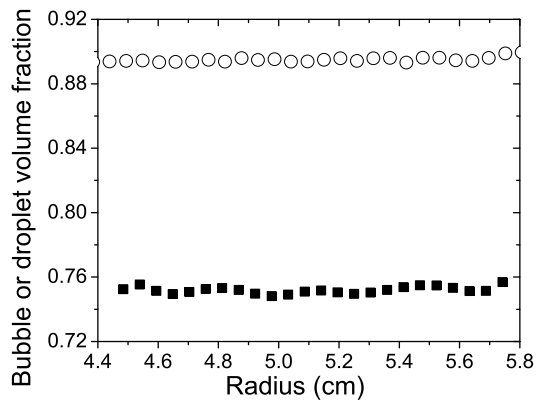}
\caption{Concentration profiles observed within the gap of a
Couette geometry by MRI techniques after shearing a concentrated
adhesive emulsion (squares, average applied strain: 70000), and a
foam (open circles, average applied strain: 12000). The emulsion
is composed of 6.5~$\mu$m diameter silicone oil droplets dispersed
at a 75\% volume fraction in a mixture of 50 wt\% glycerine and 50
wt\% water stabilized by Brij and trimethyl tetradecyl ammonium
bromide at a 6.5 wt\% concentration \cite{Ovarlez2008}. The foam
is composed of 73~$\mu$m diameter bubbles at a 89\% volume
fraction in a SLES foaming solution described in
\cite{Ovarlez2010b}.}\label{figure_volume_fraction}
\end{center}\end{figure}

Typical results are shown in Fig.~\ref{figure_volume_fraction}. We
find that
concentrated emulsions and foams remain homogeneous within
$\pm0.2\%$, even when large strains are applied; in the
experiments shown in Fig.~\ref{figure_volume_fraction}, strains
are larger than 10000. The same result is found with all emulsions
studied in \cite{Ovarlez2008}; no other foam is studied. This is in contrast to the behavior of dense suspensions of rigid particles. In particular, the strain to
reach a stationary inhomogeneous state can be as low as 20 in the suspension of rigid particles studied in
\cite{Fall2010}, of particle size similar to the droplet and bubble size of an emulsion and the foam we have studied. The absence of observable migration in concentrated emulsions
and foams may thus be due to the very different
nature of interparticle interactions in these systems: indeed,
\citet{Lhuillier2009} has recently pointed out the major role of
direct rigid contact forces in migration. Other factors may play a
role; in particular, we emphasize that the
dispersed phase volume fraction of all the soft materials studied here is well
above the jamming transition, which may hinder local increase of
the particle volume fraction.

From the absence of observable migration
for very large strains, we conclude that this process, if it
exists, would be much too slow and inefficient to lead to
observable shear banding in dense suspensions of soft particles at the
usual experiment timescales.

\vspace{0.35cm}

Finally, for nonthixotropic dense suspensions of soft
particles, thixotropy and migration should not be at the origin of shear banding. To explain the apparently contradictory
observations of the literature, one may thus wonder whether a yet
unknown mechanism is at play in some of these systems only, or if
differences in the setup or procedure are involved. In the
following, we present an experimental investigation of flow
stability in various materials with a
well-defined procedure designed to circumvent possible artefacts.

\subsection{Investigation of flow stability in dense suspensions of soft particles}\label{subsection_soft_noshbdg}

We propose to study the possible shear banding in
nonthixotropic dense suspensions of soft particles as quantified
by an intrinsic bulk property, the critical shear rate
below which flow is unstable. In order to design appropriate setup
and procedure, we first deal with possible artefacts that may lead to
observe flow inhomogeneities that are not steady-state bulk shear
bands.

\subsubsection{Possible artefacts}\label{subsubsection_soft_artefacts}

\paragraph{Finite size effects}

In systems of small gap to dispersed elements size ratio,
structural inhomogeneities are sometimes observed. Indeed, the
local shear stress\! /\! shear rate relationship $\tau(\gdot)$ in
concentrated emulsions \cite{Goyon2008,Goyon2010} and 2D foams
\cite{Katgert2010} in confined geometries is found to depend on
the system size and on the boundary conditions, suggesting
that nonlocal laws are needed to describe their behavior
\cite{Goyon2008,bocquet2009} (this is discussed in
Sec.~\ref{section_local}). The material's apparent viscosity
$\tau/\gdot$ near the walls for a given applied stress
can in particular be smaller than that of the bulk material
\cite{Goyon2008,Goyon2010}: the material is locally `fluidized' by
a rough surface. This mechanism could \textit{a priori} lead to
observe coexisting liquid-like and solid-like bands (note however
that no shear-rate discontinuity is reported in \cite{Goyon2008,Katgert2010,Goyon2010}); this would not reflect a bulk property.

\paragraph{Transient flow inhomogeneities}

\citet{Divoux2010} have recently observed long-lived transient
flow inhomogeneities during the flows of Carbopol gels in a thin
gap Couette cell. These inhomogeneities always disappear at steady
state \cite{Divoux2010}: at any rate, steady-state flow is
homogeneous. However, given the very long lifetime of these
inhomogeneities at low shear rate (they may last as long as 1 day
at a 0.1~s\m1 rate), one may wonder if all bands reported in the
literature are really steady-state bands. It is thus crucial to
understand in which condition such transient inhomogeneities are
generated; this may allow us to propose a procedure to circumvent
this possible artefact.

\begin{figure*}[htb]\begin{center}
\includegraphics[width=11cm]{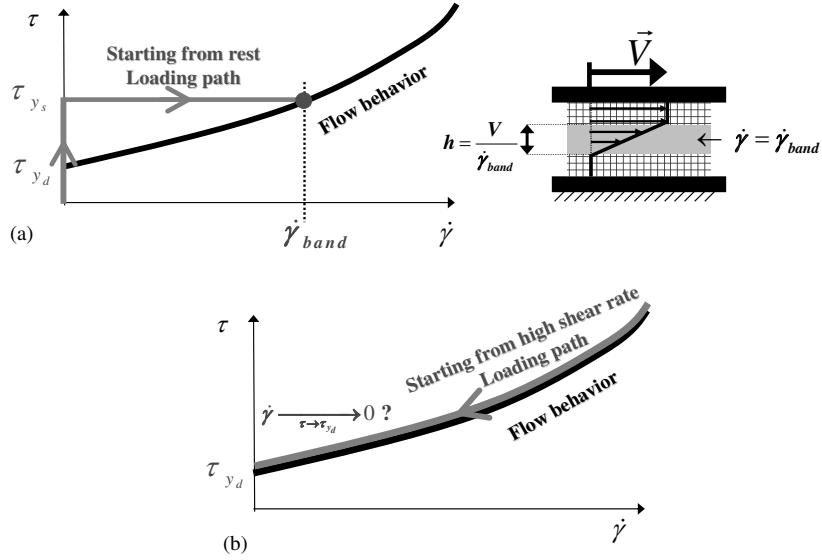}
\caption{Illustration of the procedure-dependent behavior in a
material whose static and dynamic yield stresses differ: (a) a
band may form at flow start-up
(Sec.~\ref{subsubsection_soft_artefacts}); (b) an initially
homogeneous flow may remain stable and homogeneous at any rate
when the applied shear rate is decreased
(Sec.~\ref{subsubsection_soft_noshbdg_procedure}).}\label{figure_transient}
\end{center}\end{figure*}

\noindent It is worth noting that a slight structuration at rest
is observed in the studied systems \cite{Divoux2010,Divoux2011}:
their elastic modulus increases with the resting time, and their
static and dynamic yield stress differ. Moreover, in each
experiment, flow is imposed on the material initially at rest: the
transient bands characterize flow start-up. A first basic
explanation for the development of these bands can
be given in the framework of a simple model, illustrated in
Fig.~\ref{figure_transient}a. Let us consider a material
characterized by a simple yield stress fluid behavior
$\tau=f(\gdot)$ in its liquid regime, with
$f(\gdot)\rightarrow\tau_{y_d}$ when $\gdot\rightarrow0$, which
defines its dynamic yield stress $\tau_{y_d}$. In
addition, we suppose that flow can be induced on the material
initially at rest only when $\tau\geq\tau_{y_s}$, with
$\tau_{y_s}>\tau_{y_d}$, which defines its static yield stress
$\tau_{y_s}$; this model accounts for the possibility of a
specific structuration at rest. If a macroscopic shear rate
$\gdot_{\text{macro}}$ is applied on the material initially at
rest in a slightly inhomogeneous shear stress field (as in a thin
gap Couette geometry), two cases are found, depending on the value
of $\gdot_{\text{macro}}$. If
$f(\gdot_{\text{macro}})\geq\tau_{y_s}$, after the solid/liquid
transition, the material simply flows homogeneously at the applied
shear rate $\gdot_{\text{macro}}$ and the resulting stress is
$\tau=f(\gdot_{\text{macro}})$. If
$f(\gdot_{\text{macro}})<\tau_{y_s}$
(Fig.~\ref{figure_transient}a), although the material should be
able to flow at a stress lower than the static yield stress, the
stress applied on the material in its solid regime has first to
reach $\tau_{y_s}$. Consequently, just after the solid/liquid
transition, shear must localize in a band in the region
of higher stress (near the inner cylinder in a Couette cell): the
local shear rate in the band $\gdot_{\text{band}}$ has to be
higher than the macroscopic shear rate $\gdot_{\text{macro}}$ to
ensure that the resulting stress $\tau=f(\gdot_{\text{band}})$
equals $\tau_{y_s}$, i.e.,
$\gdot_{\text{band}}=f^{-1}(\tau_{y_s})$. Note that the mechanism
proposed by \citet{Moorcroft2011} in the framework of the SGR
model is basically based on the same idea.

\noindent Of course, it is only a rough explanation of the
\citet{Divoux2010} observations\footnote{It does not explain what
happens subsequently to the band, i.e., how the material at rest
is progressively eroded by the flowing band. Moreover, the full
story of flow initiation is far more complex, as shown by
\citet{Divoux2011}; in particular, flow is initiated by full
slippage at the rotor of the Couette cell just after the
solid/liquid transition.}. However, the above simplified model
probably captures the basic reason for the existence of a banded
flow when flow is initiated on a structured state: stresses at
flow start-up have to be higher than in steady state. It also
suggests that other loading procedures can be used to avoid
generating transient bands; in the following, we propose such
procedure (Fig.~\ref{figure_transient}b).

\subsubsection{Setup and procedure}\label{subsubsection_soft_noshbdg_procedure}

We are here interested in steady-state shear bands. As defined in
Sec.~\ref{subsection_intro_shbdg}, these should result from flow
instability at low applied rate. We stress that, to investigate
their possible development, a homogeneous flow of the material in
its liquid regime should first be imposed at high shear rate;
next, the shear rate should be
decreased continuously (see Fig.~\ref{figure_transient}b). In
these conditions, transient flow inhomogeneities due to specific
structuration at rest (Sec.~\ref{subsubsection_soft_artefacts})
should not be observed: this procedure is thus preferable to
applying a given shear rate on the material initially at rest, which may lead to wrongly report shear bands
if experiments are not long enough (we recall that transient bands can last for 1 day). Then, if the material is not a
shear-banding material, homogeneous flow should be observed at any
low rate. In contrast, if the material is a shear-banding
material, flow instability should develop at low shear rate. Note that this
procedure also avoids observing the history-dependent flow
inhomogeneities recently predicted by \citet{Cheddadi2011} for
materials exhibiting significant normal stress differences; this
point is discussed in more detail below.

We focus on the behavior of dense suspensions of soft
particles: concentrated emulsions,
Carbopol gel, and foams, described in detail in
\cite{Ovarlez2008,Coussot2009,Ovarlez2010b}. To better show the
difference between these materials and thixotropic systems, we
also study the behavior of a colloidal gel (bentonite suspension),
described in \cite{Ovarlez2009}. The materials are sheared in the same
wide gap Couette geometry as in Sec.~\ref{subsubsection_soft_concentration}. This ensures that the bulk behavior, free from the finite
size effects discussed in Sec.~\ref{subsubsection_soft_artefacts}, is studied (this is shown in detail in
Sec.~\ref{section_local}). Consistently with the above
procedure, we first apply a high rotational velocity $\Omega$ to the inner cylinder of the Couette cell, and we decrease progressively
the value of $\Omega$. Steady-state azimuthal velocity profiles $V(R)$ are then measured as a function of the radial position $R$ in the gap
using MRI techniques described in \cite{Raynaud2002,Rodts2004}, for various constant rotational velocities $\Omega$.

\subsubsection{Experimental results and analysis}

\begin{figure}[htb]\begin{center}
\includegraphics[width=7cm]{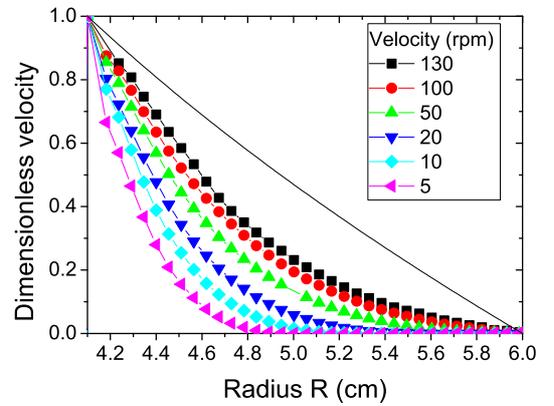}
\caption{Dimensionless velocity profiles $V(R)/V(R_i)$, measured by MRI techniques, for the
steady flows of a concentrated emulsion in a wide gap Couette
geometry, at various rotational velocities ranging from 5 to
130~rpm (see legend); the solid line is the theoretical profile
for a Newtonian fluid.
The emulsion is composed of 1~$\mu$m diameter silicone oil droplets dispersed at a 75\% volume fraction
in water
stabilized by Sodium Dodecyl Sulfate at a 8.5 wt\% concentration \cite{Ovarlez2008}.
Figure from
\citet{Ovarlez2008}.}\label{figure_exemple_profil}
\end{center}\end{figure}

All the studied materials exhibit similar behavior; an example is
shown in Fig.~\ref{figure_exemple_profil}. We observe that the
whole sample is sheared at high rotational velocity $\Omega$. When decreasing $\Omega$, below a critical value that
depends on the material and is of order 40~rpm in
Fig.~\ref{figure_exemple_profil}, the material is sheared only in
a fraction of the gap: $V(R)$ vanishes (within the measurement
uncertainty) at some radius $R_c(\Omega)$ inside the gap, which decreases
when $\Omega$ decreases.

Flow inhomogeneity here originates from the shear stress field
inhomogeneity (Eq.~\ref{eq_stress}). Shear localization should
then not be here confused with the shear banding observed in
homogeneous stress fields: it is a feature of any yield stress
fluid flow in a Couette geometry. At low $\Omega$, flow stops
at a radius $R_c$ within the gap where the local shear stress
$\tau(R)$ equals $\tau_y$ (i.e.,
$R_c=R_i\sqrt{\tau(R_i)/\tau_y}$). The decrease of $R_c(\Omega)$
as $\Omega$ is decreased then arises from the rate dependence of
the constitutive law at the approach of $\tau_y$.

Although it induces shear localization for both shear-banding and
non-shear-banding materials, the stress inhomogeneity can be used
advantageously to analyze the possible shear-banding behavior from
the shape of the localized velocity profiles. Indeed, from the
constitutive behavior point of view, the question of shear banding
can be posed as follow (Sec.~\ref{subsection_intro_shbdg}): what
is the value of the shear rate as $\tau$ tends towards the dynamic
yield stress? In shear-banding materials, it should tend towards a
nonzero value $\gdot_c$; in simple yield stress fluids, it should
tend towards zero. When flow is localized, $\tau$ approaches
$\tau_y$ when $R$ tends to $R_c$. As the local shear rate is
\bea\gdot(R)=V/R-\upd V/\upd R\label{eq_gdot}\eea the local
velocity in a shear-banding material should thus tend to zero with a
nonzero slope $|\upd V/\upd R|=\gdot_c$ at the interface between
the sheared and the unsheared regions, as in a homogeneous stress
field \cite{Ovarlez2009}. In contrast, the slope of the velocity
profile at the interface between the sheared and the unsheared
regions should be equal to zero in a non-shear-banding material.

In the above analysis, we have assumed that the yield criterion
only involves the shear stress $\tau$, whose value thus determines
the position of the interface between the sheared and the
unsheared regions. If the material displays significant normal
stress differences in its solid regime (as e.g. in dry foams
\cite{Labiausse2007}), those may enter in the yield criterion and
things can be different. Whereas steady-state normal stresses in
the flowing region depend only on the shear rate and are thus
independent of the initial conditions, different nonzero normal
stresses can be stored in the material at rest in the nonflowing
region: their value depends on shear history, i.e. on preparation.
Some sample histories may then lead to a discontinuity of the
normal stress differences at the interface between the sheared and
unsheared regions \cite{Cheddadi2011}. When taken into account in
the yield criterion, this discontinuity implies that the shear
stress at this interface is higher than in the case of continuous
normal stress differences and that $\gdot$ at this interface is
nonzero \cite{Cheddadi2011}. This leads to a slope discontinuity
in the velocity profile that appears as a critical shear rate
which is not an intrinsic property of the material, but rather
reflects the loading procedure. We point out that such
discontinuities are not expected to take place with the loading
procedure we use, as the solid-like zone is progressively formed
by decreasing continuously the inner cylinder rotational velocity.
We thus think that even with materials exhibiting significant
normal stress differences, only intrinsic shear banding will be
evidenced.

\begin{figure*}[hbt]\begin{center}
(a)\includegraphics[width=13cm]{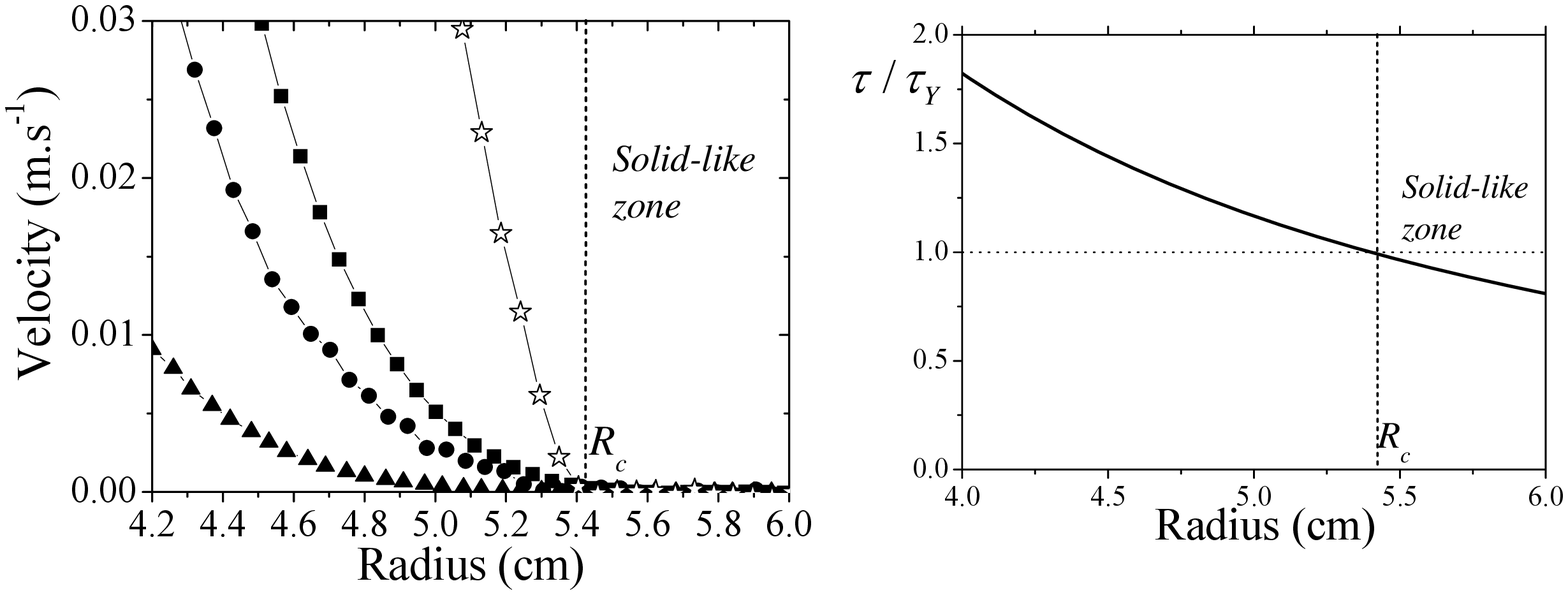}(b)
(c)\includegraphics[width=7.5cm]{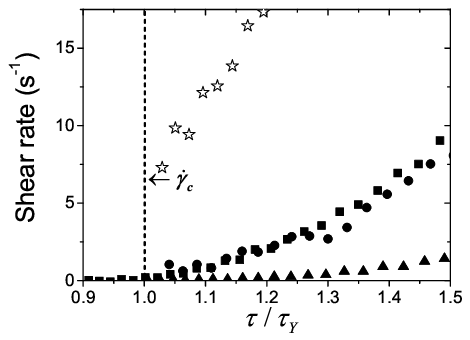}
\caption{(a) Comparison between localized velocity profiles of a
concentrated emulsion (squares), a foam (circles), a Carbopol gel
(triangles), and a bentonite suspension (empty stars). (b)
Dimensionless shear stress field $\tau(R)/\tau_y$ corresponding to
the velocity profiles of Fig.~\ref{figure_compare_profils}a. (c)
Shear rate $\gdot$ vs. shear stress extracted from the velocity
profiles of Fig.~\ref{figure_compare_profils}a. The emulsion is
composed of 1~$\mu$m diameter silicone oil droplets dispersed at a
75\% volume fraction in water stabilized by Sodium Dodecyl Sulfate
at a 8.5 wt\% concentration \cite{Ovarlez2008}. The foam is
composed of 45~$\mu$m diameter bubbles at a 92\% volume fraction
in a SLES foaming solution described in \cite{Ovarlez2010b}. The
gel is a hair gel (Vivelle Dop, France), which is mainly made up
of Carbopol in water \cite{Coussot2009}. The bentonite suspension
is made up of (smectite) clay particles of length of order
1~$\mu$m and thickness 10~nm suspended at a 4\% volume fraction in
water \cite{Coussot2010}.}\label{figure_compare_profils}
\end{center}\end{figure*}

To analyze the possible shear banding, we now plot localized
velocity profiles obtained for various materials, and
characterized by almost the same position $R_c$ (here
$\simeq5.4$~cm) of the interface between the sheared and the
unsheared regions (Fig.~\ref{figure_compare_profils}a). We focus
on the shape of the velocity profiles at this interface. From this
plot, it clearly appears that (i) the thixotropic system
(bentonite suspension) exhibits an abrupt transition from flow to
rest, as expected from the shear-banding behavior of these
materials observed in cone-and-plate geometry
\cite{Coussot2002a,Ovarlez2009}, whereas (ii) concentrated
emulsion, foam and Carbopol gel exhibit a smooth transition from
flow to rest and thus seem to be not shear-banding materials.

To better illustrate these differences and their link with the
constitutive behavior of the materials, we combine the local
stress data $\tau(R)$ (Eq.~\ref{eq_stress}) and the local shear
rate data $\gdot(R)$ (Eq.~\ref{eq_gdot}) extracted from the
localized velocity profiles
(Fig.~\ref{figure_compare_profils}a,b); this allows the
reconstruction of the local constitutive law $\tau(\gdot)$ at the
proximity of the yield stress $\tau_y$. These laws are depicted in
Fig.~\ref{figure_compare_profils}c. A clear difference between the
thixotropic system and the nonthixotropic dense suspensions of
soft particles is again observed: when $\tau\rightarrow\tau_y$, $\gdot$ tends towards a finite
value -- the critical shear rate $\gdot_c$ (here of order 6~s\m1)
-- in the bentonite suspension, and towards zero in dense
suspensions of soft particles. We have observed this smooth
transition from flow to rest, with the same
procedure, in several other emulsions \cite{Ovarlez2008} and foams
\cite{Ovarlez2010b}.

We conclude that none of the studied dense suspensions of soft
particles is a shear-banding material. Of course, due
to finite experimental resolution, we can only provide upper
bounds for the value of the critical shear rate -- if any. Note that the
accuracy of the shear rate measurement depends on the spatial
resolution and on the MRI signal to noise ratio, which depends
itself on the material and on external factors. We finally obtain
typical upper bound ranging between 0.01~s\m1 and 0.3~s\m1
\cite{Ovarlez2008,Coussot2009,Ovarlez2010b}.

\subsection{Probable explanation of previous
observations}\label{subsection_soft_literature}

From MRI observations on various systems, we claim that
nonthixotropic dense suspensions of soft particles do not show any
bulk steady-state shear banding. To strengthen this claim, we have
to check that previous observations of shear rate discontinuities
in similar systems in the literature can be explained. In the
following, we review these observations and identify possible
origins of the discrepancy between these results and ours.

\paragraph{Finite size effects} \citet{DenkovPRL2009} have reported shear banding
in 3D foams, in a parallel plate geometry of gap to bubble size
ratio of order 10. \citet{Gilbreth2006} have reported shear
banding in bubble rafts, in a geometry of gap to bubble size ratio
of order 20. Such ratios are probably much too small to ensure
that a bulk property is studied (see Sec.~\ref{section_local}). It
is worth noting that \citet{Katgert2010} reported nonlocal effect
in bubble rafts of similar gap to bubble size ratio as in
\cite{Gilbreth2006}. It is thus possible that the
\cite{Gilbreth2006,DenkovPRL2009} shear rate discontinuities are
due to finite size effects as explained in
Sec.~\ref{subsubsection_soft_artefacts}.

\paragraph{Transient effects and loading history} Transient
flow inhomogeneities may last for hours at the vicinity of the
jamming transition \cite{Divoux2010}: this casts doubt on band
observations obtained by applying a given shear rate on the
material initially at rest. Moreover, procedure-dependent steady
flow inhomogeneities are expected at flow start-up due to normal
stress differences trapped during the loading procedure in
materials which exhibit significant normal stress differences
\cite{Cheddadi2011}. Much care thus has to be taken when analyzing
inhomogeneities observed at flow start-up. In the \citet{becu2006}
experiments, shear was applied on a concentrated emulsion
initially at rest; bands were observed while the shear stress was
decreasing for 2 hours as in the \citet{Divoux2010} experiments.
We did not observe a shear-banding behavior on the same material,
when applying the procedure of
Sec.~\ref{subsubsection_soft_noshbdg_procedure} (progressive
decrease of the applied shear rate). The bands reported by
\citet{becu2006} were thus probably transient flow
inhomogeneities. The observations of \citet{Rouyer2003} on dry
foams may also find the same explanation, as these were made when
shear was applied on the material initially at rest, and as the
experiment time was rather short (the applied strain was of order
1).

\paragraph{Uncontrolled sources of thixotropy} A simple yield stress fluid can be easily
turned into a thixotropic material. E.g., \citet{Ragouilliaux2007}
have shown that tiny amounts of colloidal clay particles dispersed
in the continuous phase of a simple emulsion can form bridges
between droplets and lead to thixotropic effects and to shear
banding as detailed in Sec.~\ref{subsubsection_thixotropy}. The
same might happen with a foam. As we did not observe any shear
banding in the same foam as \citet{Rodts2005}, using the same
procedure and setup, it is suggested that the occurrence of shear
banding in their experiments might be due to uncontrolled traces
of impurities in the system, e.g. clay particles, which were also
studied with the same equipment.

\paragraph{Other artefacts} Let us finally
note that many observations of flow inhomogeneities in 2D foams
are now attributed to viscous damping at the glass boundary
\cite{Wang2006,Janiaud2006,Krishan2008,Katgert2009,Katgert2010}
when the bubbles are confined by one (or two) glass plate. Wall
drag cannot explain observations in bubble rafts with free
surface; in these last systems, as stated above, the main possible
source of such inhomogeneities stands in nonlocal effects.

\vspace{0.3cm}

All results thus seem to be consistent with the absence of
steady-state bulk shear banding in dense suspensions of soft
particles.\! Of course, all of the above explanations still have
to be checked in new experimental investigations.

\section{Consistency between macroscopic and local behavior}\label{section_local}

In this section, we address the question of the existence of a
single intrinsic local constitutive law to describe
steady-state flows of yield stress fluids. If the material can be
characterized by such law, it is expected to account for what
happens in any flow geometry, independently of the boundary
conditions. Once shear banding has been dealt with, the main source of
rheometric artefact is removed. The macroscopically measured
behavior in rheometric experiments should then reflect the local
behavior; it should thus \textit{a priori} be trusted and used to
predict flows in other configurations.

In the following, we review data from the literature where local
and macroscopic behavior do not seem to be consistent. We discuss
the main possible origins of this inconsistency: thixotropy,
volume fraction inhomogeneities, and nonlocal effects. We then
present experimental results obtained in nonthixotropic dense
suspensions of soft particles, and show that, under appropriate
conditions, in a wide gap geometry, they are characterized by a single local law that matches the
macroscopic one. We discuss the conditions needed to obtain such
result.

\subsection{Local observations in the literature}\label{subsection_local_review}

The ability of the macroscopically measured stress/strain rate
relationship to predict the flow behavior can be accurately tested
by performing local measurements of material velocity profiles
$V(\vec r)$ in geometries of controlled stress inhomogeneity
$\tau(\vec r)$ (Couette geometry, Poiseuille flow, inclined plane
flow). In such case, a theoretical velocity profile $V_{th}(\vec
r)$ is easily derived from the macroscopic relationship
$\tau_{\text{macro}}=f_{\text{macro}}(\gdot_{\text{macro}})$ as
the local shear rate should be $\gdot_{th}(\vec
r)=f^{-1}_{\text{macro}}(\tau(\vec r))$; theoretical and
experimental profiles can then be compared. An alternative
consists in deriving the local shear rate field $\gdot(\vec r)$
from the velocity profile; $\bigl(\tau(\vec r),\gdot(\vec
r)\bigr)$ data obtained at various positions $\vec r$, under
various boundary conditions, can then be combined to build the
locally derived constitutive behavior, which can be compared to
macroscopic observations.

\begin{figure}[htb]
\begin{center}
\includegraphics[width=8cm]{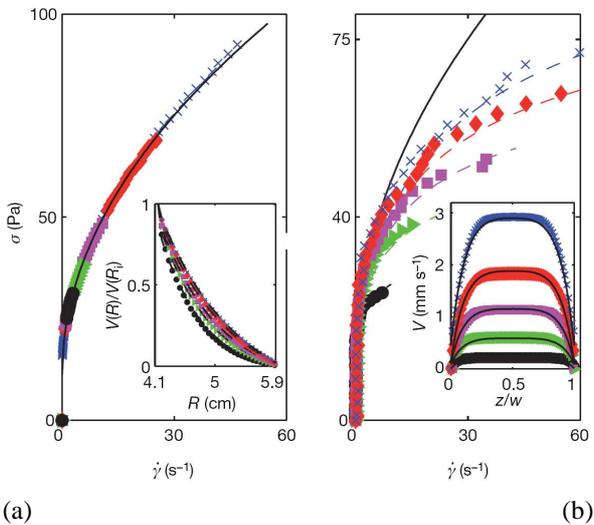}\\
\hspace{0.5cm} (a)\hfill(b)\hspace{0.5cm}\ \caption{Laws
$\tau(\gdot)$ measured locally in the gap of a 1.9~cm wide gap
Couette geometry (a) and in a $w=250~\mu$m wide microchannel (b),
during the flow of a concentrated emulsion of 75\% droplets of
6.5~$\mu$m diameter. Different symbol shapes and colors correspond
to different imposed rotational velocities of the inner cylinder
(a) or different imposed pressure drops (b). The full line is a
Herschel-Bulkley equation $\tau=\tau_y+\eta_{_{HB}}\gdot^n$, with
$n=0.5$, $\tau_y=11.6$Pa and $\eta_{_{HB}}=11.2$Pa.s$^{n}$. The
insets show the velocity profiles from which these laws are
extracted. The lines in Fig.~\ref{figure_goyon}b inset are fits of
the velocity profiles to the nonlocal behavior
Eq.~\ref{eq_nonlocal}. Figure from
\citet{Goyon2008}.}\label{figure_goyon}
\end{center}
\end{figure}

Such local measurements in concentrated emulsions have
yielded surprising results
\cite{Salmon2003,becu2006,Goyon2008}: in some emulsions, the
locally measured constitutive law does not match the
macroscopically measured one. Moreover, it seems impossible to
find a single local constitutive law $\tau(\gdot)$ compatible with
all flows. The apparent $\tau(\gdot)$ law needed to describe the
velocity profile for a given boundary condition does not match
those needed for other boundary conditions: it
depends on the velocity at the inner cylinder in a Couette cell
\cite{Salmon2003,becu2006} or the pressure gradient in a
Poiseuille cell \cite{Goyon2008} (see an example in
Fig.~\ref{figure_goyon}b). The same features have been observed in
dense noncolloidal suspensions \cite{Huang2005} and 2D foams
(bubble rafts) \cite{Katgert2010} sheared in Couette
geometries.

Three main reasons can been mentioned to explain the apparent
absence of a single local constitutive law: transient effects,
shear-induced migration of the dispersed elements, and nonlocal
effects.

\subsubsection{Impact of thixotropy}\label{subsubsection_local_thixo}

\begin{figure}[htb]
\begin{center}
\includegraphics[width=7.5cm]{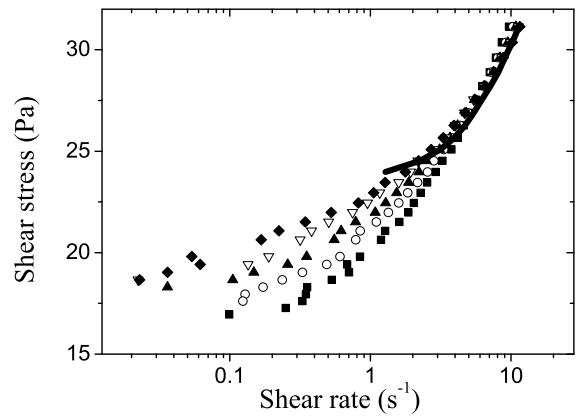}\\
\caption{Laws $\tau(\gdot,t)$ measured locally at various times
$t$ (squares: 30~s, empty circles: 90~s, up triangles: 150~s,
empty down triangle: 210~s, diamonds: 270~s, full line: 1~h)
during the flow of a thixotropic mud (see \cite{Ragouilliaux2007}
for details on the material) in a wide gap Couette geometry. The material
is first sheared at a rotational velocity $\Omega=50$~rpm during
10~min; the velocity is then decreased abruptly at
$\Omega=15$~rpm and 2 velocity profiles are recorded per minute
during 1~h, which allows reconstructing the time-dependent
local law $\tau(\gdot,t)$.}\label{figure_ragouilliaux}
\end{center}
\end{figure}

In a thixotropic material, the shear
stress $\tau$ is a function of the shear rate $\gdot$ and of
shear history. A single relationship between $\tau$ and $\gdot$
can be defined only in steady state. During a transient, the shear
rate can take many values for a given applied stress $\tau$,
depending on time and on shear history. When local
measurements are not strictly performed in steady state, it may
then leave the false idea that different laws are required to
describe the flow of the sample. This is illustrated in
Fig.~\ref{figure_ragouilliaux} where local laws $\tau(\gdot,t)$
measured in a thixotropic material at various times $t$ during a
transient are displayed. It is clearly observed that the local
flow behavior depends on shear history, until a steady-state is
reached.

Although dense suspensions of soft particles are usually found to
be nonthixotropic, it has been shown that long times
can be needed to reach a steady state at flow start-up in some of
these systems \cite{Divoux2010} (Sec.~\ref{subsubsection_soft_artefacts}). This seems to be due to a
specific structuration mechanism at rest; the structure formed at
rest may then take time to be erased at low shear rate. Two
situations should thus be distinguished: (i) when the
material is at rest, it gets slightly structured and its behavior
at flow start-up may first depend on flow history during long time
(as long as 24~h in the \citet{Divoux2010} experiments); (ii)
when the material is flowing at steady state, its subsequent flow behavior has no significant dependence on
flow history as long as it is continuously strained in its liquid
state. It contrasts with usual thixotropic systems such as
colloidal gels, where competition between structuration and
destructuration is observed in the liquid state.

These structuration effects might explain the observations of
\citet{becu2006}, who cannot account for the flows of an
adhesive emulsion with a single constitutive law. Their
experiments are indeed performed at flow start-up. Their
observation of flow inhomogeneity, whereas we do not observe shear banding -- on the same material -- when slowly decreasing the
applied shear rate from high to low values (see previous section),
then suggests that their local measurements are performed during a
transient, consistently with the \citet{Divoux2010} observations.

\subsubsection{Impact of particle migration}\label{subsubsection_local_migration}

Particle migration can also be a cause for the apparent lack of a
local constitutive law \cite{Ovarlez2006}. Indeed, when various
$\bigl(\tau(\vec r),\gdot(\vec r)\bigr)$ data are combined to
build a constitutive law, a hidden hypothesis is that the material
is homogeneous, i.e., that the same material is dealt with at any
position $\vec r$ in the flow. If there is an inhomogeneous volume
fraction field $\phi(\vec r)$ in the material, one has to be
careful when inferring a constitutive law from velocity profiles.
To illustrate the problem, let us consider a volume fraction
dependent constitutive behavior $\tau=g(\phi,\gdot)$. Under two
different boundary conditions denoted $b_1$ and $b_2$, a same
local value $\gdot_0$ of the shear rate can be reached at two
positions $\vec r_{b_1}$ and $\vec r_{b_2}$ where the
local $\phi$ values differ. When reconstructing a
constitutive law from all local data, this may lead to associating
two different stress values $\tau_{b_1}=g(\phi(\vec
r_{b_1}),\gdot_0)$ and $\tau_{b_2}=g(\phi(\vec r_{b_2}),\gdot_0)$
to the same shear rate $\gdot_0$, thus leading to the -- erroneous
-- conclusion that flow under different boundary conditions is not
compatible with a single constitutive behavior.

This phenomenon explains the apparent lack of a
single constitutive law to account for all flows in dense
noncolloidal suspensions of rigid particles \cite{Huang2005}.
Further studies of the same system have indeed shown that it is
inhomogeneous \cite{Ovarlez2006} due to shear-induced particle
migration. A local constitutive behavior compatible with all flows
can nevertheless be built \cite{Ovarlez2006,Fall2010} by combining
$\bigl(\tau(\phi(\vec r_0),\gdot(\vec r_0)),\gdot(\phi(\vec
r_0),\gdot(\vec r_0))\bigr)$ data measured at a fixed position
$\vec r_0$ under various boundary conditions: this provides the
constitutive behavior at a well defined volume fraction $\phi(\vec
r_0)$ for various local shear rates $\gdot(\vec r_0)$. By studying
this local behavior at various positions $\vec r_0$, a
$\phi$-dependent local behavior is finally built. The observations
of flow inhomogeneity, and thus of apparent viscosity
inhomogeneity, in a colloidal glass sheared in a homogeneous
stress field \cite{besseling2010} is also accounted for by
volume fraction inhomogeneities.

This phenomenon is not expected to occur in dense suspensions of
soft particles: we have shown in
Sec.~\ref{subsubsection_soft_concentration}
that these systems remain homogeneous under shear. Consequently,
it probably cannot explain the \citet{Goyon2008} and
\citet{Katgert2010} observations.

\subsubsection{Nonlocal behavior}\label{subsubsection_local_nonlocal}

The apparent absence of a single local constitutive law can also
be the signature of nonlocal phenomena. Indeed, the local
$\tau(\gdot)$ relationship observed in dense emulsions in microchannels by
\citet{Goyon2008} was found to depend on the boundary conditions
(pressure drop, surfaces roughness), which is the hallmark of
nonlocal behavior (Fig.~\ref{figure_goyon}b).

A nonlocal modeling \cite{Goyon2008,bocquet2009} has been shown
to successfully account for all flows in the
\citet{Goyon2008,Goyon2010} experiments; fair agreement has also
been found with the \citet{Katgert2010} experiments in bubble
rafts. In jammed materials, flow basically occurs locally via a
succession of elastic deformation and irreversible plastic events.
The main idea of the nonlocal model \cite{Picard2005} is that a
localized plastic event induces relaxation of the stress in the
whole system, and may thus significantly affect the behavior in
its neighborhood -- its zone of influence being quantified by a
length $\xi$, called `flow cooperativity length'. Consequently,
the rate of plastic rearrangements in a given zone depends on the
rate of rearrangements in its neighborhood. It will then be
different in a strongly inhomogeneous stress field, i.e., if the
stress varies significantly over $\xi$, than in a homogeneous
stress field far from boundaries (in this last situation, the bulk
behavior is obtained). Nonlocal effects are thus predicted to take
place in the presence of high stress gradients. In its simplest
form, the \citet{bocquet2009} model is
\bea[\gdot/\tau](z)=[\gdot/\tau]_{bulk}(z)+\xi^2\partial^2_z[\gdot/\tau](z)\label{eq_nonlocal}\eea
where $[\gdot/\tau]_{bulk}$ is the bulk apparent inverse viscosity
of the material (also called `fluidity') and $z$ is the direction
of the stress gradient \cite{Goyon2008}. This nonlocal model
also accounts for the impact of the wall roughness, when the
material displays a different behavior near the boundary than in
the bulk \cite{Goyon2008,Goyon2010}. In a
homogeneous stress field, the length over which the behavior
changes from surface to bulk behavior is then predicted to be
of the order of a few $\xi$.

These two features explain the \citet{Goyon2008,Goyon2010}
observations in microchannels of small size. From a fit
of experimental data to the nonlocal model
(Fig.~\ref{figure_goyon}b inset), they have shown
that $\xi$ is zero below the jamming volume fraction $\phi_m$, and
increases continuously with $\phi$ above $\phi_m$. E.g., $\xi$ is
found to be 5 droplet diameters in a monodisperse emulsion of 85\%
volume fraction. Consistently, discrepancy between the local and
macroscopic behavior is found in 10 to 40 droplets wide
geometries but not in a 3000 droplets wide geometry
\cite{Goyon2008} (see
Fig.~\ref{figure_goyon}). $\xi$ is found to be of the order of 3
bubble diameters in the bidisperse bubble raft studied by
\citet{Katgert2010}, which is 20 bubbles wide; no large scale
study is reported for rafts.

\vspace{0.35cm}

In nonthixotropic dense suspensions of soft particles, the effects
discussed above can \textit{a priori} be dealt with. Indeed,
migration is not an issue
(Sec.~\ref{subsubsection_soft_concentration}) and significant
time-dependent behavior may be found only in specific conditions
(Sec.~\ref{subsubsection_soft_artefacts}); moreover, the impact of
nonlocal effects should be minimal in wide gap geometries. In the
following, we propose to study the local behavior of such
materials in well-defined conditions where the above mechanisms
should not be at play.

\subsection{\!Nonthixotropic dense suspensions of soft particles in a wide gap
geometry}\label{subsection_local_soft}

Our goal is to see if, under some conditions, a simple local law
accounts for yield stress fluid flows and matches their
macroscopic response. Given the effects discussed above,
appropriate setup and procedures must be used.

\subsubsection{Setup and procedure}\label{subsubsection_local_soft_procedure}

In order to ensure that the bulk behavior, free from the finite
size (nonlocal) effects discussed in
Sec.~\ref{subsubsection_local_nonlocal}, is studied, the flow
geometry that is used should have specific characteristics. The
nonlocal model (Eq.~\ref{eq_nonlocal}) and the observations of
\citet{Goyon2008} first imply that the gap has to be at least tens
of cooperativity length $\xi$ wide, i.e., more than one hundred
dispersed elements diameters wide to minimize the impact of the
boundaries. Moreover, from Eq.~\ref{eq_nonlocal}, smooth stress
spatial variations (of a maximum order of a few \% over $\xi$) are
required.

Here, we use the same wide gap Couette geometry as in
Secs.~\ref{subsubsection_soft_concentration}
\&~\ref{subsection_soft_noshbdg}. In a Couette geometry, the
stress is inhomogeneous (Eq.~\ref{eq_stress}) and the stress
relative variation over a length $\xi<<R_i$, where $R_i$ is the
inner cylinder radius, is
$|\tau(R_i+\xi)/\tau(R_i)-1|\simeq2\xi/R_i$. Assuming a value of
$\xi$ of the order of 5 dispersed elements diameter $d$, as found
by \citet{Goyon2008} in some systems, smooth stress spatial
variations impose $R_i$ to be at least equal to several hundreds
of $d$. Note that this is only a rough estimate as these are the
variations of the `fluidity' $\gdot/\tau$ that have to be smooth.
We thus have limited our studies to materials whose dispersed
elements diameter is less than 100 $\mu$m to meet the above
requirements (with $R_i=4.1$~cm and gap$=1.9$~cm).

To ensure that steady-state properties are studied, the long-lived
transient effects discussed in
Sec.~\ref{subsubsection_soft_artefacts} and their possible
consequence on the local behavior
(Sec.~\ref{subsubsection_local_thixo}) should be avoided. That is
why we use the same procedure as in
Sec.~\ref{subsubsection_soft_noshbdg_procedure}, i.e., we first
impose a homogeneous flow of the material at high shear rate,
before decreasing progressively the applied shear rate on the
flowing -- liquid-like -- material (see
Fig.~\ref{figure_transient}b).

The same materials as in Sec.~\ref{subsection_soft_noshbdg} are
studied: concentrated emulsions, Carbopol gel, and foams,
described in detail in
\cite{Ovarlez2008,Coussot2009,Ovarlez2010b}. After a 5 min
preshear at 50 to 100 rpm (depending on the material), we measure
both the steady-state azimuthal velocity profiles $V(\Omega,R)$
with MRI techniques (see
Sec.~\ref{subsubsection_local_soft_procedure}) and the torque
$T(\Omega)$ with a Bohlin C-VOR 200 rheometer, for various
constant rotational velocities $\Omega$ of the Couette cell inner
cylinder. In all systems, the stationary velocity profiles are
found to develop within a few seconds and to remain stable for
long durations (as long as hours in the emulsions), which means
that these systems are not thixotropic \cite{Coussot2005}. For the
\citet{becu2006} emulsion, this contrasts with the long time
evolution of the velocity profiles they observe. We argue that
this comes from the differences in the procedures used, as
discussed in Sec.~\ref{subsubsection_soft_noshbdg_procedure}; this
validates the procedure we use to study a behavior free from
transient effects.

\subsubsection{Experimental results and analysis}\label{subsubsection_local_soft_results}

\begin{figure}[htb] \begin{center}
\includegraphics[width=8.5cm]{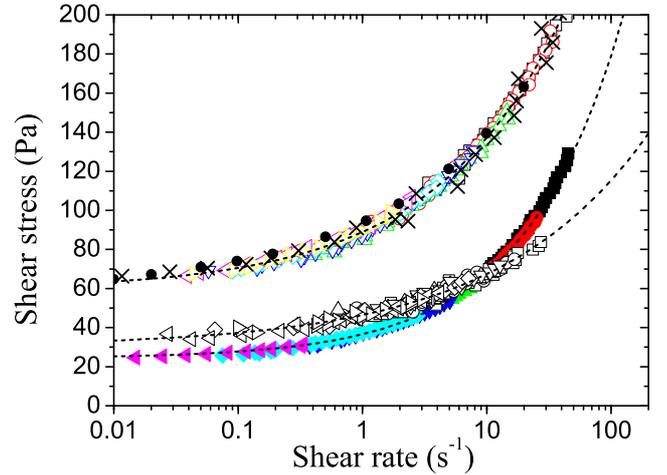}
\caption{Constitutive laws of (from bottom to top) a concentrated
emulsion, a foam, and a Carbopol gel, measured locally in a wide
gap Couette cell using MRI techniques. Macroscopic data measured
in the Carbopol gel in the Couette geometry (crosses) and in a
cone and plate geometry (filled circles) are also shown. The
dotted lines are Herschel-Bulkley fits to the data
$\tau=\tau_y+\eta_{_{HB}}\gdot^{n}$ with: $\tau_y=24.3$~Pa,
$\eta_{_{HB}}$=12.3~Pa\,s$^{0.55}$, and $n=0.55$ (emulsion);
$\tau_y=30.3$~Pa, $\eta_{_{HB}}$=16.2~Pa\,s$^{0.36}$, and $n=0.36$
(foam); $\tau_y=59.3$~Pa, $\eta_{_{HB}}$=29.6~Pa\,s$^{0.42}$, and
$n=0.42$ (Carbopol gel). The emulsion is composed of 6.5~$\mu$m
diameter silicone oil droplets dispersed at a 75\% volume fraction
in a mixture of 50 wt\% glycerine and 50 wt\% water stabilized by
Brij and trimethyl tetradecyl ammonium bromide at a 1 wt\%
concentration \cite{Ovarlez2008}. The foam is composed of
45~$\mu$m diameter bubbles at a 92\% volume fraction in a SLES
foaming solution described in \cite{Ovarlez2010b}. The gel is a
hair gel (Vivelle Dop, France), which is mainly made up of
Carbopol in water
\cite{Coussot2009}.}\label{fig_compare_constitutive_laws}
\end{center}\end{figure}

Typical steady-state velocity profiles are shown in
Fig.~\ref{figure_exemple_profil}. The constitutive laws of the
materials accounting for their velocity profiles can be built by
using both the velocity profiles and the torque measurements. The
stress distribution $\tau(R)$ at a radial position $R$ within the
gap is obtained from Eq.~\ref{eq_stress}; we recall that
$\tau(R_i)=T/(2\pi R_i^2H)$ where $H$ is the height of the inner
cylinder. The local shear rate $\gdot(R)$ in the gap is inferred
from the velocity profiles $V(R)$ thanks to Eq.~\ref{eq_gdot}.
Both measurements performed at a given radius $R$ for a given
rotational velocity $\Omega$ provide a local data point of the
constitutive law $\tau=f(\gdot)$. Note that, as
Eqs.~\ref{eq_stress} \&~\ref{eq_gdot} are valid whatever the
boundary conditions are, this analysis is not affected by a
possible wall slip. As the materials are homogeneous, we are
allowed to combine the $\bigl(\tau(R),\gdot(R)\bigr)$ data
measured at various positions $R$ (see
Sec.~\ref{subsubsection_local_migration}): they are \textit{a
priori} representative of the same material. The local
constitutive laws extracted from the experiments performed at
various rotational velocities $\Omega$ on an emulsion, a foam, and
a Carbopol gel, are plotted in
Fig.~\ref{fig_compare_constitutive_laws}.

A first important result is that, for each material, all the shear
stress vs. shear rate data fall along a single curve. This means
that, for a given material, data obtained under different boundary
conditions (different $\Omega$) effectively reflect the behavior
of a single material with a given local intrinsic constitutive law
in simple shear. This observation contrasts with the \citet{Goyon2008} observations, on the same emulsion, consistently with the
predictions of the nonlocal model discussed in
Sec.~\ref{subsubsection_local_migration}:
nonlocal effects are not expected in the geometry we use as the
gap is 3000 droplets wide for this material (thus ensuring that both surface effects
and stress inhomogeneities have no observable impact on the
behavior). We also contrast our observations on 3D foams with
those on rafts by \citet{Katgert2010}; again, this is consistent with what can be expected from
the nonlocal model as the gap size is 400 bubbles wide in our
experiments, and 20 bubbles wide in the \citet{Katgert2010}
experiments. Our measurements performed on the \citet{becu2006}
emulsion contrast with their observations too. In the
\citet{becu2006} study, however, the gap being 1mm for 0.3$\mu$m
droplets, the nonlocal effects discussed in
Sec.~\ref{subsubsection_local_nonlocal}
cannot be the reason for the apparent absence of a single
constitutive law accounting for all flows. As already pointed out
above, we think that the main difference between the two
experiments stands in the procedure as the \citet{becu2006}
experiments were performed at shear start-up on a structured
material. This tends to show that, provided an appropriate
procedure is used, a single local behavior is recovered.

It is worth noting that data are obtained at local shear rates as
low as $10^{-2}$~s\m1, the lower measurable value depending on the
resolution on the velocity measurement (thus on its
derivative), and also on the coarsening rate for foams
\cite{Hohler2005}. This observation is consistent with the absence
of observable shear banding (Sec.~\ref{subsection_soft_noshbdg}). All observed behaviors are
finally consistent with a continuous local law $\tau(\gdot)$ with
$\gdot\rightarrow0$ when $\tau\rightarrow\tau_y$. Good agreement
is found with a Herschel-Bulkley equation
$\tau(\gdot)=\tau_y+\eta_{_{HB}}\gdot^n$ for all materials in a
[$10^{-2}$~s\m1, 40~s\m1] range of shear rates; the fitting
parameters (in particular the index $n$) depend on the material,
and are given in the figure caption.

We now get back to the macroscopic behavior. The torque data
$T(\Omega)$ obtained in the Couette geometry can be analyzed using
standard rheometric analysis (without using the velocity profiles)
to yield the macroscopic constitutive law
$\tau=f_{\text{macro}}(\gdot)$ of the material; this analysis has
to be performed carefully to account for shear localization
\cite{Coussot2005,Ovarlez2008,Coussot2009} when computing the
macroscopic shear rate from the value of the inner cylinder
rotational velocity. These macroscopic data, displayed in
Fig.~\ref{fig_compare_constitutive_laws} for the
Carbopol gel, are in good agreement with the local data. Good
agreement has also been found in the case of an emulsion
\cite{Ovarlez2008}. To push the analysis further, we have also
characterized the Carbopol gel with a 40~mm diameter 4\!\degre angle
sandblasted cone-and-plate geometry. In such geometry, the stress
field is homogeneous and cannot be a source of nonlocal effects;
however, the gap still has to be wide enough to measure bulk
properties. In the used geometry, the gap varies between 1.4~mm
at the edges and 150~$\mu$m (truncature height); the size of the
dispersed elements of the studied material should thus be smaller
than $\approx20\mu$m. The macroscopic data obtained in the
Carbopol gel with this geometry are shown in
Fig.~\ref{fig_compare_constitutive_laws}. Good agreement is
obtained with the macroscopic and local data measured in the
Couette geometry. This shows that a single intrinsic local
behavior indeed characterizes the flows of this material. This
validates the use of macroscopic tools to study the behavior of
these systems, provided the gap is wide enough and a procedure
similar to the one we propose (shearing continuously from high to
low shear rate) is used.

\vspace{0.35cm}

Altogether, these results show that, in the strict
conditions discussed in
Sec.~\ref{subsubsection_local_soft_procedure}, a simple yield
stress fluid behavior exists: at steady-state, all the studied
nonthixotropic dense suspensions of soft particles are indeed
characterized by a single, intrinsic, local, continuous and monotonic
constitutive law in simple shear. For all materials, it is
moreover consistent with a Herschel-Bulkley equation, as usually
found in standard rheological experiments.

\section{Conclusion}

Although many complex behaviors have been observed in yield stress
fluid flows, we have shown that, under well-defined conditions, a
class of materials, namely nonthixotropic dense suspensions of
soft particles, display a simple yield stress fluid behavior: at
steady-state, they are characterized by a single, intrinsic,
local, continuous monotonic constitutive equation $\tau(\gdot)$ in
simple shear with $\gdot\rightarrow0$ when $\tau\rightarrow\tau_y$
(i.e., they are not shear-banding materials).

\vspace{0.3cm}

To obtain these results, the main material characteristics are
that:
\begin{itemize}
\item they have to be nonthixotropic. More precisely, they should
not display time-dependent behavior in their liquid state,
otherwise they will show shear banding. Possible structuration at
rest is not an issue if an appropriate characterization procedure
is used (see below).\item they should be made up of soft
particles. Dense suspensions of soft particles have been shown to
remain homogeneous when sheared, in contrast to suspensions of
rigid particles.
\end{itemize}
Materials complying with these requirements are foams,
concentrated emulsions, and Carbopol gels.

\vspace{0.3cm}

As regards the setup and procedure,
two points have to be respected:
\begin{itemize}
\item the geometry where flow is studied should have a wide gap,
and should be characterized by smooth stress inhomogeneities to
ensure that nonlocal effects are not present. We recall that the
relevant lengthscale to consider in this problem may be a
`cooperativity length' rather than the dispersed elements size, as
found in concentrated emulsions. \item the following procedure
should be applied: flow has first to be imposed at high shear rate
to erase the memory of possible structuration at rest; afterwards,
the material flow should be studied by progressively decreasing
the applied shear rate, without passing by a resting period. This
ensures that possible transient shear banding is avoided and that
the local flow behavior is consistent with the macroscopic one.
\end{itemize}

In these conditions, a simple yield stress fluid behavior is
finally shown to exist. For the studied materials, the steady flow
behavior in simple shear can be modeled by a Herschel-Bulkley
equation, as found in standard rheological experiments. A
macroscopic characterization can then be trusted as it reflects
the local intrinsic behavior of the material. It can \textit{a
priori} be used to predict the flow behavior in other
configurations, provided the studied flow also complies with the
above requirements (large flow scale, problem different from flow
start-up at low shear rate).

Of course, when confined flows or slow start-up flows are studied,
all the complex effects discussed throughout the text will
possibly occur; they thus still have to be studied in depth in
fundamental studies to help predicting what happens in such
situations. In some materials, other effects such as elastic
effects should also be incorporated in the modeling as they may be
of crucial importance in 3D flows.

\section*{Acknowledgment}
We acknowledge financial support from the European Space Agency (Contract No. MAP AO 99-108: C14914/02/NL/SH).

\end{document}